# Photonic spin lattices: symmetry constraints for skyrmion and meron topologies

Xinrui Lei[1,2,†], Aiping Yang[1,†], Peng Shi[1], Zhenwei Xie[1], Luping Du[1,*], Anatoly V. Zayats[2,*], Xiaocong Yuan[1,*]

[1]Nanophotonics Research Centre, Shenzhen Key Laboratory of Micro-Scale Optical Information Technology, Shenzhen University, 518060, China

[2]Department of Physics and London Centre for Nanotechnology, King's College London, Strand, London, WC2R 2LS, United Kingdom

**Abstract:**
Symmetry governs many electronic and photonic phenomena in optics and condensed matter physics. Skyrmions and merons are prominent topological structures in magnetic materials, with the topological features determined by the interplay between anisotropy of a material and its magnetization. Here we theoretically show and experimentally demonstrate that the symmetry of the electromagnetic field determines the spin topological properties of the guided modes via spin-orbit coupling and may only result in either hexagonal spin-skyrmion or square spin-meron lattices. We also show that in the absence of spin-orbit coupling these spin topologies are degenerated in dynamic field-skyrmions, unifying description of electromagnetic field topologies. The results provide new understanding of electromagnetic field topology and its transformations as well as new opportunities for applications in quantum optics, spin-optics and topological photonics.

**Main:**
Topologically nontrivial spin textures are of great significance in a various field of physics, ranging from high-energy to condensed matter physics[1-5]. Magnetic skyrmions and merons are topologically protected nanoparticle-like objects with a magnetization swirl in chiral or non-centrosymmetric magnets, important for practical applications in spintronics[6-10]. Depending on the properties of ferromagnetic materials and/or external magnetic fields, skyrmions, having integer topological number, can exist as isolated entities or condense into a lattice. In contrast to skyrmions, merons are topological constructs with a half-integer topological charge. Because of this, the individual merons cannot be observed and magnetic merons always form a lattice[10,11]. Recently, photonic analogies of magnetic skyrmions were developed with either optical spin forming a skyrmion-like topology (photonic spin-skyrmions)[12] or a lattice of the electric field textures with hedgehog structure (photonic field-skyrmions, which are dynamic field textures with the electric field oscillating in time)[13,14]. In the former case, spin-orbit coupling[15] in the evanescent field provides a topological protection through the generalized photonic quantum spin-Hall effect[16], leading to the locking of the spin to the local momentum of the evanescent wave[17]. In the latter case, the symmetry of the electromagnetic field imposed due to the interference in the 6-fold symmetry cavity, results in the creation of the field-skyrmions in a hexagonal lattice. Photonic skyrmions are considered for applications in nanophotonics, quantum technologies, metrology and high-resolution imaging[12,14,18].

Here we show that it is the symmetry of the field that completely determines the topology of the electromagnetic field of the guided modes (the modes with the evanescent field component) through spin-orbit coupling. We demonstrate both experimentally and theoretically the spin-skyrmion and spin-meron lattices formed due to a broken rotational symmetry of the field, with 6-fold symmetry being responsible for a skyrmion lattice and 4-fold symmetry for a meron lattice. In the absence of the spin-orbit coupling (in the case when the field is not carrying topological charge), instead of spin-skyrmions, field skyrmions are formed, thus connecting two types of skyrmion manifestation. We also show that the spin skyrmion and spin meron textures correspond to the lowest energy of the electromagnetic field configuration, therefore, energetically stable. The demonstrated photonic spin-topologies open up new pathways for topological photonics, quantum photonics and metrology and new avenue to explore topological condensed matter systems.

Individual photonic spin-skyrmions formed due to the spin-orbit coupling in the guided waves with the evanescent field component (Fig. 1a) can be described by a Hertz vector potential with a helical phase term in the cylindrical coordinate $(r, \varphi, z)$ as[19]

$$\Psi_0 = AJ_l(k_r r)e^{il\varphi}e^{-k_z z}, \quad (1)$$

where $A$ is a constant, $k_r$ and $ik_z$ are the transverse and longitudinal wave-vector components satisfying $k_r^2 - k_z^2 = k^2$ with $k$ being the wave-vector, and $J_l$ is the Bessel function of the first kind of order $l$. The Neel-type photonic spin-skyrmions are formed for $l \neq 0$, with the spin up or down in the center of the skyrmion, determined by the sign of $l$. Equation (1) describes an evanescent optical vortex (eOV) in a source free, homogeneous and isotropic medium. If the rotational symmetry of the field is broken, individual eOVs will interact and may condense in a lattice. In order to reveal the topological features under different types of symmetry, we consider superpositions of the individual Hertz vector potentials with each lattice point fed by an eOV: $\Psi = \Psi_0 * \sum \delta(\mathbf{r} - \mathbf{r}_{mn})$, where $\mathbf{r}_{mn}$ is the position of each lattice point, $\delta$ is the Dirac delta function and * represents the operation of convolution. In two dimensions, only two Bravais lattices exist with equal lattice constants corresponding to point groups $D_6$ and $D_4$ having hexagonal and square lattices, respectively (Figs. 1a and e). In view of translational and rotational symmetry imposed by a lattice, the total Hertz potential can be obtained as (see Supplementary Information for the details of the derivations)

$$\Psi = A\sum_{n=1}^{2N} e^{il\theta_n}e^{ik_r \mathbf{r}\cdot\mathbf{e}_n}e^{-k_z z}, \quad (2)$$

where $\theta_n = n\pi/N$, $\mathbf{e}_n = (\cos\theta_n, \sin\theta_n)$ with $N=3$ for hexagonal and $N=2$ for square lattice.

The Hertz potential in Equation (2) exhibits rotational periodicity with total angular momentum $l$ (we consider $l = 1$ in the following). The amplitude distributions of the Hertz potential manifest the lattice symmetry features (Figs. 1a and e) with zero points corresponding the singularities of the phase distribution (Figs. 1b and f).

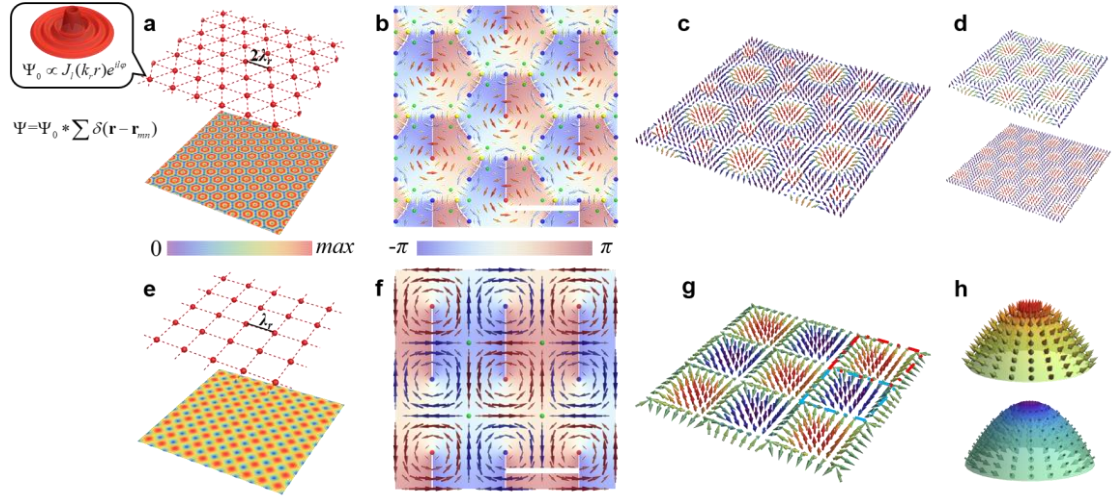

**Fig. 1 | Formation of spin-skyrmion and spin-meron lattices via spin-orbit coupling in the fields of different symmetries. a,** Schematic of the eOV lattice of hexagonal symmetry with a lattice constant $2\lambda_r$ ($\lambda_r$ is in-plane wavelength) and the distribution of the amplitudes of the resultant Hertz potential. Each lattice point (red dots) is fed with an eOV described by a helical Hertz potential $\Psi_0$ corresponding to isolated spin-skyrmion (insert). **b,** Poynting vector direction (arrows) and magnitude (arrow colors) in the generated optical vortex lattice in **a**. Background color represents the phase distribution of total Hertz potential. Red and yellow dots denote the center of the anticlockwise Poynting vector vortices (due to two sub-lattices), where the phase of the Hertz potential is singular; blue dots denote the center of clockwise ones. Green dots show the junction of different Poynting vector vortices where $P=0$ with nonzero Hertz potential. The scale bar is $\lambda_r$. Note that thin vertical line in each vortex is an artifact of the colour representation where phase is the same on both sides. **c,** Optical spin orientation distribution corresponding to the lattice in **a-b** showing a spin-skyrmion lattice. The arrows indicate the direction of the unit spin vector. **d,** Two sub-lattices of the skyrmion lattice forming the spin distribution in **c**. Central "up" state in each sub-set can be referred to red and yellow dots in **b**, respectively. **e,** Schematic and the distribution of the Hertz potential amplitude for the evanescent optical vortex lattice of square symmetry with a lattice constant $\lambda_r$. **f,** Same as **b** for the square lattice in **e**. The scale bar is $\lambda_r/2$. **g,** Same as **c** for the square lattice in **e-f**. The dashed red and blue squares indicate unit cells of the spin-meron lattice with skyrmion number $Q = \pm \frac{1}{2}$, which can be smoothly mapped on a paraboloid in **h**.

The spin texture of the electromagnetic field can be obtained considering the Poynting vector distribution in the lattice since for TM/TE polarized evanescent waves, the intrinsic spin-momentum coupling yields a relationship between the SAM **S** and the Poynting vector **P** as[17]

$$\mathbf{S} = \frac{1}{2\omega^2} \nabla \times \mathbf{P}, \qquad (3)$$

where $\omega$ is the angular frequency of the wave. The generalized spin momentum relationship in Equation (3) is a manifestation of the conservation law of SAM, which indicates that the SAM of an evanescent field originates from the vortices of the electromagnetic energy flow. The Poynting vector $\mathbf{P}=\text{Re}(\mathbf{E}^*\times\mathbf{H})/2$ represents a directional energy flux of an electromagnetic field in the lattice, which can be calculated through the Hertz potential as $\mathbf{P}=-\frac{\omega\varepsilon k_r^2}{2}\text{Im}(\Psi\nabla\Psi^*)$ [20], where $\varepsilon$ is the absolute permittivity of the medium (See Supplementary Information for details). The phase singularity of the Hertz potential pins the center of each Poynting vector vortex, and depending on the symmetry of the lattice, the in-plane Poynting vector forms different types of vortex distributions.

For a hexagonal lattice, the Poynting vector possesses a rotational self-similarity with characteristic angle of $\pi/3$ as $\hat{R}(\frac{\pi}{3})\mathbf{P}(\mathbf{r})=\mathbf{P}[\hat{R}(\frac{\pi}{3})\mathbf{r}]$, where $\hat{R}(\theta)$ is the rotation matrix along $z$-axis. A hierarchical structure of the Poynting vector vortices is obtained with two sublattices, due to the multiple vortex centers in each unit cell of the lattice (Fig. 1b), which can be considered as an optical analogue of the Abrikosov vortex lattice[21]. The appearance of two sub-sets of the Poynting vector vortex lattice can be attributed to the different compositions of the wave numbers (see Supplementary Information for details). While for a square lattice, the Poynting vector possesses a $\pi/2$ angled rotational self-similarity as $\hat{R}(\frac{\pi}{2})\mathbf{P}(\mathbf{r})=\mathbf{P}[\hat{R}(\frac{\pi}{2})\mathbf{r}]$. The energy flow in each unit cell exhibits a vortex with winding number $\pm 1$ (Fig. 1f), which is an optical analogue of the so-called staggered-flux in condensed matter systems[22,23]. The junction of the vortices in Figs. 1b and f can be found where the Poynting vector is zero but the Hertz potential is non-zero.

These two types of vortices in the Poynting vector distributions determine the topology of the spin textures. The topological invariant $Q$, which is known as the skyrmion number, can be obtained as $Q=(1/4\pi)\iint \mathbf{n}\cdot(\partial_x\mathbf{n}\times\partial_y\mathbf{n})dxdy$, where $\mathbf{n}=\mathbf{S}/|\mathbf{S}|$ represents the unit vector in the direction of a three-component spin. Since each vortex center of the Poynting vector is the phase singularity of the Herz vector, where the Poynting vector is zero, only $z$-component of the SAM is present in the center of the vortex (Equation (3)), with sign of $S_z$ determined by the rotation direction of the Poynting vector vortex. For a hexagonal lattice, two skyrmion sub-lattices are observed, according to the structure of the Poynting vector distribution, with skyrmion number $Q = 1$ for each unit cell of the sublattice where the spin vectors vary progressively from the central "up" state to the edge "down" state, manifesting a Neel-type photonic spin-skyrmion (Figs. 1c and d).

On the other hand, the staggered-flux configuration of the Poynting vector observed in a square lattice gives rise to the formation of a photonics spin-meron lattice. From Equation (3), the SAM distribution can be obtained as $\mathbf{S} \propto (k_z\sin(k_rx)\cos(k_ry), k_z\cos(k_rx)\sin(k_ry), k_r\cos(k_rx)\cos(k_ry))$, which possesses distinct domains where $z$-component of the local spin orientation goes to zero and merons are confined in a unit cell (Fig. 1g). In each unit cell, the spin vector tilts progressively from the central 'up' or 'down' state to the edge where $z$ component of SAM $S_z = 0$. The skyrmion number calculated for each unit cell in the spin texture is $Q = \pm \frac{1}{2}$, corresponding to the Neel-type photonic 'core-up' or 'core-down' spin-meron topology. These spin configurations in each unit cell can be smoothly mapped on a paraboloid $z = -r^2/2$ by converting the in-plane spin vector to an infinite circle with substitution $\frac{k_z}{k_r}\tan(k_r x) = r\cos\varphi$ and $\frac{k_z}{k_r}\tan(k_r y) = r\sin\varphi$ (Fig. 1h). The obtained square spin-meron lattice with alternating Poynting vector vorticities is in analogy with the lowest-energy multi-meron configuration in frustrated magnets[24,25].

It is worth noting that in the case of $l = 0$, the SAM vanishes and, in the absence of spin-orbit coupling, the spin skyrmion and meron are not present. In this case, the considered 6-fold and 4-fold symmetry lattices of the Herz potential result in the oscillating electric field patterns which exhibit field-topologies corresponding to a field-skyrmion lattice for D6 and a field-meron lattice for D4 symmetry (see Supplementary Information and Fig. S1 for details). This provides connection between the concepts of spin-skyrmions formed due to the spin-orbit coupling[12] and dynamic, time-dependent field-skyrmions formed due to the interference effects[13,14].

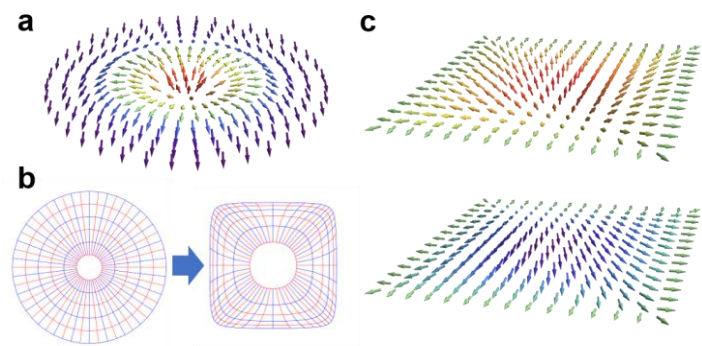

**Fig. 2 | Formation of spin meron lattice via boundary conversion. a,** Isolated skyrmion configuration in a cylindrical system derived from lowest energy theorem in the nonlinear sigma model. **b,** Coordinate transformation from an infinite circle to a confined square when a square spin lattice is formed. **c,** Two meron configurations confined in a square corresponding to two solutions of Equation (4) with $Q = \pm \frac{1}{2}$, forming the spin-lattice in Fig. 1g.

The configuration of a square meron lattice can be understood from the thermodynamic perspective considering the lowest-energy theorem. For magnetic skyrmions, the energy Hamiltonian is given by the nonlinear sigma model together with the Dzyaloshinskii-Moriya (DM) interactions for the antisymmetric exchange between two neighboring magnetic spins[8,26]. A meron lattice can be stabilized in chiral magnets with anisotropy[27,28]. Considering the Hamiltonian in the classical nonlinear sigma model[29]

$$H = \frac{1}{2}\int (\nabla \mathbf{n})^2 dxdy, \qquad (4)$$

the individual spin-skyrmion solution can be obtained in the system with the rotational symmetry as $\theta(r)=2\arctan(r)$ by using Euler–Lagrange (EL) equation for the lowest energy (see Supplementary Information for details) and applying boundary conditions $n_z(0)=1$ at the core and $n_z(\infty)=-1$ at infinity (Fig. 2a). If we chose the boundary conditions corresponding to a square spin lattice: $n_z(0)=\pm 1$ at the core and $n_z(\infty)=0$ at the edge of the unit cell (Fig. 2b), the exact numerical solution for a meron state can be obtained behaving as $\theta(r)\approx\arctan(r)$ or $\theta(r)\approx\pi/2-\arctan(r)$ (Fig. 2c). These solutions correspond to the SAM configuration of a square lattice in Fig. 1g (see Supplementary Information for details). It was not possible to find an analytical solution to Equation (4) because of the difficulties with defining boundary conditions due to the composite sub-lattice structure.

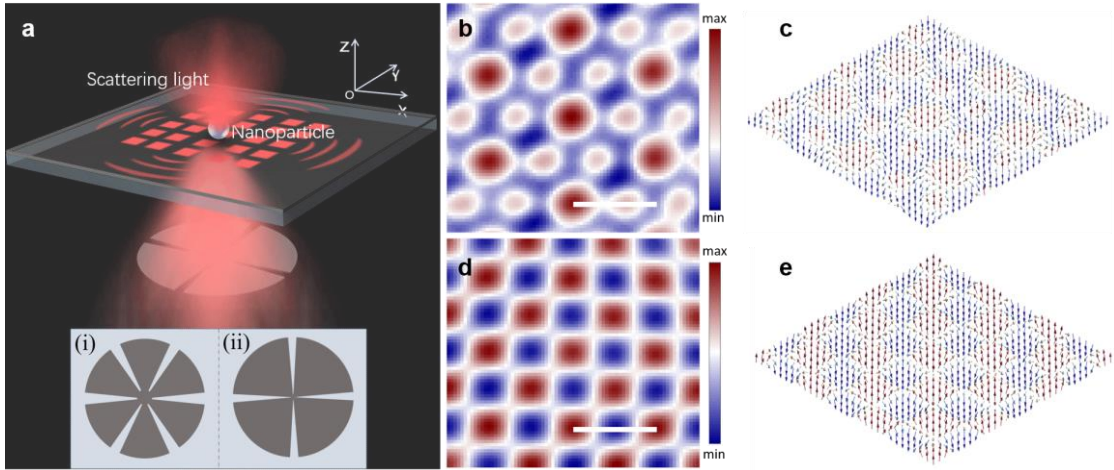

**Fig. 3 | Experimental demonstration of photonic spin-skyrmion and spin-meron lattices in the SPP field with spin-orbit coupling. a,** Schematic diagram of the experiment. SPPs are excited on a surface of a thin silver film by a tightly focused, radially polarized beam modulated by the intensity masks comprised of six- or four-fold symmetry apertures. A dielectric nanosphere is employed as a near-field probe to characterize the spin-texture of the spin lattice, formed at the center area induced by the spin-orbit interaction in SPPs. **b-e,** Measured $S_z$ component **(b,d)** and local spin orientation **(c,e)** in the lattice obtained with the field of six-fold **(b-c)** and four-fold **(d-e)** symmetry. The scale bar in **b** and **d** is the SPP wavelength. In **c** and **e**, the arrows indicate the orientation of the normalized spin vectors, with the $z$-direction colour-coded to the same colour scale as **b** and **d**.

We experimentally demonstrated the spin lattices of the evanescent vortex beams on the example of surface plasmon polaritons (SPPs) sustained at a dielectric-metal interface[30]. A radially polarized laser beam with helical wavefront (topological charge $l=1$, wavelength $\lambda=532$ nm) is tightly focused onto a thin silver film surface, which provides an excitation of SPPs with OAM at air/silver interface. Under such conditions, individual photonic spin-skyrmions are formed on the metal interface in the case if rotational symmetry is not broken[12]. To break the rotational symmetry, the incident beam was modulated with intensity masks comprised of either six- or four-fold symmetry apertures. The spin-skyrmion or spin-meron textures are formed in the case of six-fold or four-fold symmetry, respectively, due to interference of the SPPs in the presence of the spin-orbit interaction (Fig. 3a). A spin-resolved near-field scanning optical microscope was used with a dielectric nanosphere as a near-field probe to characterize the spin-texture of the vortex lattice[31] (see Supplementary Information for the details of the experiment). The measured $S_z$ distributions and the retrieved local spin vector orientation exhibit a skyrmion-type lattice for the field of six-fold symmetry and reveal spin-meron lattice for four-fold symmetry (Figs. 3b-e). These distributions correspond well to the simulation results for respective symmetries of the plasmonic field (cf. Figs. 3c,e and 1c,g), revealing to interleaved hexagonal skyrmion lattices for a hexagonal lattice and square lattice of positive/negative merons.

In conclusion, we demonstrated the formation of photonic spin-skyrmion and meron lattices in real-space based on spin-orbit coupling in the evanescent waves in the environment with broken rotational symmetry. We showed that the field symmetry is the key for determining the photonic spin-topology, resulting either 6-fold symmetry spin-skyrmion lattices or 4-fold symmetry of spin-meron lattices. We experimentally verified these two kinds of spin-lattice topologies on the example of SPP waves in the presence of spin-orbit coupling in the broken rotational symmetry conditions. In the absence of spin-orbit coupling, the spin lattices are degenerated to time-dependent field-skyrmion or field meron lattices. Thermodynamic considerations confirm stability of spin-textures of the evanescence waves. These new topological features of electromagnetic waves may provide new insights on the properties of skyrmion and meron topological structures and their transformations in condensed matter physics, where they may be difficult to realize, as well as new applications in topological and quantum photonics and spin optics.

**Methods:** See the Supplementary Information for further methods.

**Data availability:** All data is available in the main text and Supplementary Information.

**Acknowledgements:** This research was supported by the Guangdong Major Project of Basic Research No. 2020B0301030009, National Natural Science Foundation of China grants U1701661, 61935013, 62075139, 61622504, 61705135 and 61905163, the leadership of Guangdong province program grant 00201505, the Natural Science Foundation of Guangdong Province grant 2016A030312010, the Shenzhen Peacock Plan KQTD20170330110444030 and KQTD2015071016560101, the Science and Technology Innovation Commission of Shenzhen grants JCYJ20200109114018750, the EPSRC (UK) and the ERC iCOMM project (789340). L.D. acknowledges the support given by the Guangdong Special Support Program.

**Author contributions:** L. D. developed the concept of the work. X. L., P. S., Z. X. and L. D. carried out the analytical and numerical modeling. L. D. designed the experiment. A. Y. performed the experiments. X. L., L. D., A.V. Z. and X. Y. analyzed the data. X. L., L. D. and A.V. Z. wrote the manuscript. L. D., A.V. Z. and X.Y. supervised the entire project. All authors discussed the results and commented on the article.

**Competing interests:** Authors declare no competing interests.

**Additional information:**
**Supplementary information** Supplementary Text and Figures S1-S5.
**Correspondence and requests for materials** should be addressed to L. D., A.V. Z. or X.Y.